\documentstyle[aps,epsfig]{revtex}   

\begin{document}
\title{The particle in the box: Intermode traces in the propagator}
\author{I. Marzoli, I. Bialynicki-Birula\footnote{Permanent address:
        Center for Theoretical Physics, Lotnik\'ow 46, 02--668 Warsaw,
        Poland}, O. M. Friesch, A.~E.~Kaplan\footnote{Permanent 
        address: Electr. \& Comp. Eng. Dept., The Johns Hopkins 
        University, MD--21210, USA} and 
        W.~P. Schleich}
\address{Abteilung f\"{u}r Quantenphysik, Universit\"{a}t Ulm,
89069 Ulm, Germany}
\maketitle
\begin{abstract}
Characteristic structures such as canals and 
ridges---intermode traces---emerge in the
spacetime representation of the probability distribution of a
particle in a one-dimensional box.
We show that the corresponding propagator already contains 
these structures.
We relate their visibility to the factorization property of the initial
wave packet.
\end{abstract}
\section{Introduction}
The Zeeman or Stark effect in a hydrogen atom \cite{bohm}, the phase 
change at a conical 
intersection in a polyatomic molecule \cite{herzberg} and the
nuclear shell structure \cite{mottelson} are three striking phenomena
illustrating the importance of degeneracies in quantum mechanics.
But how can degeneracies play a key role in such a well studied 
problem as the particle in the box?
In the present paper we show that the characteristic
structures~\cite{kinzel,berry,stifter,marklof,grossmann,marzoli} 
in the spacetime representation of a particle in a box
shown in Figs.~\ref{zero} and \ref{combo} and referred to as 
quantum carpet are a consequence of a manifold degeneracy~\cite{kaplan}.

Our analysis relies on three observations:
\begin{enumerate}
\item    The wave function of the particle at time $t$ is the integral
of the initial wave function and the
well known propagator of the box problem~\cite{bohm}.
\item The probability of finding the particle is the
absolute value squared of the wave function. 
In the propagator formulation of quantum mechanics this translates 
into the two-dimensional integral of the product
of the initial wave function and its complex conjugate multiplied by
a kernel.   This kernel is the
product of the Green's function and its complex conjugate evaluated 
at the two different initial conditions.
\item  The quadratic dependence of the energy on the quantum number
factorizes in the product of the Green's functions into the product 
of two linear dependences.
\end{enumerate}
These observations allow us to rewrite the kernel in a way 
which brings out
most clearly that the carpet structure is inherent to the propagator.
Three results stand out:
\begin{enumerate}
\item
 The kernel is only non-vanishing along straight lines in spacetime.
\item The steepness of these lines is discrete.
\item These lines either start at the difference or the sum of 
the integration variables.
\end{enumerate}
But where is the degeneracy?
The first degeneracy has its roots in the dispersion relation connecting
in a quadratic way kinetic energy and wave number.
This forces 
many quantum numbers to contribute to the same spacetime line.
Another degeneracy arises when the product of the initial 
wave function times its 
complex conjugate factorizes into the product of two functions  which 
now contain the sum and the difference of the integration 
variables, only. 
This degeneracy in the initial wave function either enhances or 
suppresses the intermode traces resulting from the spectrum degeneracy.
We note that the phenomenon of fractional 
revivals~\cite{berry,aronstein,stifter1} is 
closely related to the intermode traces.

Our paper is organized as follows.
In Sec.~\ref{s2} we cast the probability to find the particle at time
$t$ at the position $x$ in the box into a kernel formulation.
Within Sec.~\ref{s3} we rewrite this kernel to bring out the intermode
traces and discuss in Sec.~\ref{s4} the consequences of the factorization
property for the emergence of the carpet.
We illustrate these results in Sec.~\ref{s5} for the example of a 
Gaussian wave packet.
We conclude in Sec.~\ref{s6} by summarizing our main results and
giving an outlook.
\section{The particle in the box: a brief summary}
\label{s2}
In this section we briefly summarize the essential ingredients 
of the problem of the particle in the box.
In particular, we express the probability $W(x,t)$
to find the particle at time $t$ at position $x$ as an integral of
the product of the initial wave packet, its complex conjugate 
and a kernel.
This kernel is the product of the two Green's functions at two different
initial positions.

We consider the motion of a particle of mass $M$ caught between two
walls separated by a length $L$.
The wave function $\psi(x,t)$ of this particle at time $t$ reads
\begin{equation}
\psi(x,t) = \sum_{m=1}^{\infty} \psi_m \, u_m(x) 
\exp\left(-i E_m t /\hbar \right) \,.
\label{wave_func}
\end{equation}
Here the expansion coefficients 
\begin{equation}
\psi_m \equiv \int_0^L dx' \, u_m(x') \, \varphi(x')
\label{expansion}
\end{equation}
follow from the wave function $\psi(x,t=0) \equiv \varphi(x)$ at
time $t=0$ and from the energy eigenfunctions
\begin{equation}
u_m(x) \equiv {\cal N} \sin(k_m x) \equiv \sqrt{ \frac{2}{L} } \,
       \sin \left( m \pi \frac{x}{L} \right) 
\label{eigenstates}
\end{equation}
with eigenvalue
\begin{equation}
E_m \equiv \frac{ (\hbar k_m)^2}{2M} = m^2 \hbar \frac{2\pi}{T}\,.
\label{eigenvalues}
\end{equation}
Here we have introduced the revival time $T\equiv 4ML^2/(\pi\hbar)$.

Note that the form of the energy eigenfunctions
 $u_m$ together 
with the wave numbers
$k_m$  guarantee that the wave function $\psi$ satisfies the boundary
conditions
\begin{equation}
\psi(x=0,t) = \psi(x=L,t) = 0
\end{equation}
enforced by the two walls at $x=0$ and $x=L$ at all times.

The constant ${\cal N} \equiv (2/L)^{1/2}$ results from the 
normalization condition
\begin{equation}
1 = \int_0^L dx \, u_m(x)^2 = {\cal N}^2 \int_0^L dx \,
\frac{1}{2} \left[ 1 - \cos(2 k_m x) \right] = {\cal N}^2 \frac{L}{2} \,.
\end{equation}

For further calculations it is more convenient to have the summation
in the representation Eq.~(\ref{wave_func}) of the wave function run 
from minus infinity to plus infinity rather than from unity to 
plus infinity.
For this purpose we first decompose the energy 
eigenfunctions
\begin{equation}
u_m(x) = \sqrt{\frac{1}{2L}} \frac{1}{i} \left( 
e^{im\pi x/L} - e^{-im\pi x/L} \right)
\end{equation}
into right and left running waves which yields
\begin{equation}
\psi(x,t) = \frac{1}{i\sqrt{2L}} \sum_{m=1}^{\infty} \left\{
           \psi_m \exp\left[ i\pi  m \left( \frac{x}{L}
           - m\frac{2t}{T} \right) \right] 
           -  \psi_m \exp\left[ -i\pi  m \left( \frac{x}{L}
           + m\frac{2t}{T} \right) \right] \right\} \,.
\end{equation}
When we define the expansion coefficients $\psi_m$ for negative 
values of $m$ by
\begin{equation}
\psi_{-|m|} \equiv - \psi_{|m|}
\end{equation}
we find the compact representation
\begin{equation}
\psi(x,t) = \frac{1}{i\sqrt{2L}} \sum_{m=-\infty}^{\infty} \psi_m
            \exp\left[ i\pi  m \left( \frac{x}{L} - m \frac{2t}{T}
            \right) \right] 
\label{compact}
\end{equation}
of the wave function.
Here we have used the fact that $\psi_0 = 0$ since $u_0 = 0$.

When we now substitute the explicit form
\begin{equation}
\psi_m = \frac{1}{i\sqrt{2L}} \int_0^L dx' \, \left[
         \exp\left( i\pi  m \frac{x'}{L} \right)
       - \exp\left(-i\pi  m \frac{x'}{L} \right) \right] \varphi(x')
\end{equation}
of the expansion coefficients into the expression Eq.~(\ref{compact})
of the wave function we arrive at
\begin{equation}
\psi(x,t) = \int_0^L dx' \, \varphi(x') \, G(x,t\,|\,x') \,. 
\end{equation}
Here we have introduced the Green's function
\begin{equation}
G(x,t\,|\,x') \equiv \frac{1}{2L} \sum_{m=-\infty}^{\infty} \left[
          \exp\left(-i\pi  m \frac{x'}{L} \right) 
         -\exp\left( i\pi m \frac{x'}{L} \right) \right] 
          \times \exp\left[ i\pi  m \left( 
          \frac{x}{L} - m\frac{2t}{T} \right) \right]
\label{green}
\end{equation}
of the particle in the box.

With the help of this result we can represent the probability 
\begin{equation}
W(x,t) = |\psi(x,t)|^2 = \int_0^L dx' \int_0^L dx'' \, \varphi^*(x') \,
         \varphi(x'') \, K(x,t\,|\,x',x'')
\label{probab_kernel}
\end{equation}
to find the particle at time $t$ at the position $x$
in terms of the initial wave function $\varphi$, its complex
conjugate $\varphi^*$ and the kernel
\begin{equation}
K(x,t\,|\,x',x'') \equiv G^*(x,t\,|\,x') \cdot G(x,t\,|\,x'')
\label{kernel}
\end{equation}
consisting of the product of the two Green's functions at the two 
different initial positions.

In the next section we show that the kernel already contains
the intermode traces in the carpet.
They emerge in a striking way provided the initial wave packet satisfies
an appropriate factorization condition.
\section{A new representation of the kernel}
\label{s3}
In this section we first 
find a new representation of the kernel.
We then discuss its properties.  
For the details of the calculations we refer to the appendix.

We first calculate the kernel defined in Eq.~(\ref{kernel}).
For this purpose 
we substitute the expression Eq.~(\ref{green}) for the Green's function
$G$ and an analogous expression for
$G^*$ into the definition, Eq.~(\ref{kernel}), of the kernel. 
After minor algebra we arrive at
\begin{eqnarray}
K(x,t\, | \,x',x'') &=& 
      \frac{1}{4L^2} \sum_{m',m''=-\infty}^{\infty} \exp\left\{
      -i\pi  (m'-m'') \left[ \frac{x}{L} - (m'+m'') \frac{2t}{T}
      \right] \right\} \nonumber \\
&\times& \left\{ \exp\left[ i\pi  \left(m' \frac{x'}{L} 
                            -m''\frac{x''}{L} \right) \right]
       +         \exp\left[-i\pi  \left(m' \frac{x'}{L}
                            -m''\frac{x''}{L} \right) \right] \right.
\nonumber \\
&&- \left. \exp\left[ i\pi  \left(m' \frac{x'}{L}
                     +m''\frac{x''}{L} \right) \right]
       -   \exp\left[-i\pi  \left(m' \frac{x'}{L}
                     +m''\frac{x''}{L} \right) \right] \right\} \,.
\end{eqnarray}
When we recall the relations
\begin{equation}
m' x' - m'' x'' = (m'-m'') \, \frac{x'+x''}{2} 
                + (m'+m'') \, \frac{x'-x''}{2}
\end{equation}
and
\begin{equation}
m' x' + m'' x'' = (m'+m'') \, \frac{x'+x''}{2} 
                + (m'-m'') \, \frac{x'-x''}{2}
\end{equation}
we can cast the kernel into the form  
\begin{eqnarray}
K(x,t \, | \, x', x'') = 
      \frac{1}{4L^2} && \left[ {\cal D}\left( \frac{x'-x''}{2L} ;
      -\frac{x'+x''}{2L} \right) 
      + {\cal D}\left( -\frac{x'-x''}{2L} ; \frac{x'+x''}{2L} \right)
      \right. \nonumber \\
&&-\left. {\cal D}\left(\frac{x'+x''}{2L} ;-\frac{x'-x''}{2L} \right)
 -        {\cal D}\left(-\frac{x'+x''}{2L} ;\frac{x'-x''}{2L} \right)
\right] \,.
\label{kernel1}
\end{eqnarray}
It consists of four contributions, each of which involves the function
\begin{equation}
{\cal D}(\eta ; \zeta) \equiv \sum_{m',m''=-\infty}^{\infty}
\exp \left[ i \pi (m'+m'') \eta \right] \,
\exp \left\{ -i \pi (m'-m'') \left[ \frac{x}{L} - (m'+m'')\frac{2t}{T}
+ \zeta \right] \right\} \,.
\end{equation}  
Note that the four terms differ in the arguments of ${\cal D}$.

The function ${\cal D}$ looks rather complicated.
However in the appendix we derive the equivalent representation
\begin{equation}
{\cal D}(\eta ; \zeta) = \sum_{l,n=-\infty}^{\infty} (-1)^{n\cdot l} \;
e^{i\pi n \eta} \; \delta[ \chi_{n,l} (x,t) + \zeta ]
\end{equation}
where
\begin{equation}
\chi_{n,l} (x,t) \equiv \frac{x}{L} - n \frac{t}{T/2} -l \,.
\label{chi}
\end{equation}
With the help of this relation the kernel $K$ for the probability
$W$ to find the particle at the time $t$ at position $x$ reads
\begin{eqnarray}
K(x,t \, | \, x',x'') = \frac{1}{4L^2}  && \left\{ 
  \sum_{l,n=-\infty}^{\infty} (-1)^{n\cdot l} \exp\left(i\pi n
  \frac{x'-x''}{2L}\right) \, \delta\left[ \chi_{n,l}(x,t) 
  -\frac{x'+x''}{2L} \right] \right. \nonumber \\
&&+  \sum_{l,n=-\infty}^{\infty} (-1)^{n\cdot l} \exp\left(-i\pi n
  \frac{x'-x''}{2L}\right) \, \delta\left[ \chi_{n,l}(x,t)
  +\frac{x'+x''}{2L} \right]         \nonumber\\
&&-  \sum_{l,n=-\infty}^{\infty} (-1)^{n\cdot l} \exp\left(i\pi n
  \frac{x'+x''}{2L}\right) \, \delta\left[ \chi_{n,l}(x,t)
  - \frac{x'-x''}{2L} \right]         \nonumber\\
&&-\left. \sum_{l,n=-\infty}^{\infty} (-1)^{n\cdot l} \exp\left(-i\pi n
  \frac{x'+x''}{2L}\right) \, \delta\left[ \chi_{n,l}(x,t)
  + \frac{x'-x''}{2L} \right] \right\} \,.
\label{kernel2}
\end{eqnarray}

This is the key result of the present paper.
We now discuss the properties of the kernel using this representation.
Here we focus in particular on the behavior of $K$ in the spacetime 
strip defined by the left and the right walls at $x=0$ and $x=L$.

We note that the kernel $K$ has quite a characteristic form.
Due to the appearance of the $\delta$-functions it is different
from zero only along the four sets of straight spacetime lines
\begin{equation}
\frac{t_{n,l}(x;x'+x'')}{T/2} \equiv \frac{1}{n} \left( \frac{x}{L}
\mp \frac{x'+x''}{2L} -l \right)
\label{lines1}
\end{equation}
and
\begin{equation}
\frac{t_{n,l}(x;x'-x'')}{T/2} \equiv \frac{1}{n} \left( \frac{x}{L}
\mp \frac{x'-x''}{2L} - l \right) \,.
\label{lines2}
\end{equation}
The steepness $n^{-1}$ of these world lines is quantized in terms of the 
integer $n$, which due to the summation can take on any value 
from minus infinity to plus infinity.

These lines enter or leave the spacetime strip through the left wall at
$x=0$ at the times
\begin{equation}
\frac{t_{n,l}(x=0;x'+x'')}{T/2} = \mp\frac{1}{n} \frac{x'+x''}{2L}
- \frac{l}{n}
\end{equation}
and 
\begin{equation}
\frac{t_{n,l}(x=0;x'-x'')}{T/2} = \mp\frac{1}{n} \frac{x'-x''}{2L}
- \frac{l}{n} \,.
\end{equation}
We note that apart from the integers $n$ and $l$ the integration 
variables $x'$ and $x''$ determine
these times.
Hence for one set of $n$ and $l$ values these spacetime lines are 
determined by the integration variables $x'$ and $x''$. 
However, there is a degeneracy in these lines: 
the integration variables $x'$ and $x''$ 
enter these expressions {\it either\/} in their sum {\it or\/} their
difference.
Hence it is left to the
initial wave packet to either enforce or suppress this degeneracy.
This either enhances or reduces the visibility of the intermode 
traces, as we show in the next section. 

We emphasize that there is already one degeneracy guaranteed by the
quadratic energy spectrum Eq.~(\ref{eigenvalues}) of the particle in
the box: In the function ${\cal D}$ the summation indices $m'$ and
$m''$ enter the expression in the square brackets multiplying time
in their sums $m'+m''$ only.
Indeed by multiplying it with the prefactor $m'-m''$ outside of
the brackets we realize that this is a consequence of the quadratic
dispersion relation connecting kinetic energy and wave number.
This sum $m'+m''$ leads eventually to the integer value $n$ determining
the steepness of the spacetime lines defined in Eq.~(\ref{chi}).
Hence many values of the summation indices $m'$ and $m''$ 
lead to the same steepness $n^{-1}$.

According to Eq.~(\ref{kernel2}) each $\delta$-function
has a complex valued prefactor.
This prefactor contains the term $(-1)^{n\cdot l}$ as well as the
exponential $\exp[\pm i\pi n (x'-x'')/(2L) ]$ or 
$\exp[\pm i\pi n (x'+x'')/(2L) ]$.
Again as in the argument of the $\delta$-function 
{\it either\/} the sum {\it or\/} the difference of the integration 
variables $x'$ and $x''$ enters.
Hence there is an interesting factorization property in $K$:
When the difference $x'-x''$ occurs in the argument of the 
$\delta$-functions the phase of the complex amplitude 
contains the sum $x'+x''$ and vice versa.
This factorization property is essential for the emergence of the 
intermode traces as we discuss in Sec.~\ref{s4}.

We conclude this section by noting that the kernel $K$ for the 
probability to find the particle at time $t$ at the position $x$ in the
box consists of $\delta$-functions aligned along straight 
spacetime lines.
These world lines have discrete steepness and enter or leave the 
spacetime strip at
$x=0$ at times determined by the sum and the difference of the 
integration variables $x'$ and $x''$ of the initial wave packet.
\section{Emergence of intermode traces}
\label{s4}
In the preceding section we have realized that the kernel factorizes
into sums of products of two functions.
Each of the functions either depends on the sum $x'+x''$ or the
difference $x'-x''$ of the integration
variables $x'$ and $x''$, that is, of the position variables of the
initial wave packet $\varphi^*(x')$ and $\varphi(x'')$. 
We are therefore lead to consider wave packets  which satisfy
the factorization property
\begin{equation}
\varphi^*(x') \, \varphi(x'') = \varphi_+ (x'+x'') \, \varphi_-(x'-x'')
\label{fact_property}
\end{equation}
where $\varphi_+$ and $\varphi_-$ are new wave functions.
Moreover, the initial wave packet has to satisfy the boundary conditions.
Are there any wave functions that factorize and vanish at the walls?
Here we do not want to go into a full discussion of this question
but only mention that a narrow Gaussian satisfies both requirements in 
an approximated sense as discussed in Sec.~\ref{s5}.

For the case of factorizable wave functions we can now perform one of the
two integrations with the help of the $\delta$-functions in the kernel
$K$.
Then the probability to find the particle at time $t$
at position $x$ reads
\begin{equation}
W(x,t) = \int_0^L dx' \int_0^L dx'' \, \varphi_-(x'-x'') \,
\varphi_+(x'+x'') \, K(x,t \, | \, x', x'') \,.
\label{W}
\end{equation}
Under the assumption of a very narrow 
initial wave packet, it is possible to extend the integration limits
to minus infinity and plus infinity.
Motivated by the factorization property of the kernel discussed in the
preceding section we now introduce the
more convenient integration variables
\begin{equation}
\xi_+ \equiv \frac{1}{2L} \, (x'+x'') \,, 
\end{equation}
and 
\begin{equation}
\xi_- \equiv \frac{1}{2L} \, (x'-x'') \,.
\end{equation}
We make use of the expression, Eq.~(\ref{kernel2}), for the kernel $K$
and the probability distribution becomes
\begin{eqnarray}
W(x,t) &=& \frac{1}{2} \sum_{l,n=-\infty}^{\infty} (-1)^{n\cdot l}
\int_{-\infty}^{\infty} d\xi_  - \int_{-\infty}^{\infty} d\xi_+ \,
\varphi_-(2L\xi_-) \, \varphi_+(2L\xi_+) \nonumber \\
&\times& \left\{ e^{i\pi n \xi_-} \, \delta\left[ \chi_{n,l}(x,t) -
\xi_+ \right] + e^{-i\pi n \xi_-} \, 
\delta\left[ \chi_{n,l}(x,t) + \xi_+ \right] \right. \nonumber\\
&&- \left. e^{i\pi n \xi_+} \, \delta\left[ \chi_{n,l}(x,t) -
\xi_- \right] - e^{-i\pi n \xi_+} \, 
\delta\left[ \chi_{n,l}(x,t) + \xi_- \right] \right\} \,.
\end{eqnarray}
The presence of $\delta$-functions allows us to integrate over one
variable.   
The remaining integral is the Fourier transform
\begin{equation}
\tilde{\varphi}_\pm(\kappa_n) \equiv \frac{1}{2L} \int_{-\infty}^{\infty}
dy \, e^{i\kappa_n y} \, \varphi_\pm (y) \,
\end{equation}
of the factorized wave packets $\varphi_-$ and $\varphi_+$
to the wave number variable
$\kappa_n \equiv n\pi/(2L)$.
Hence we find
\begin{eqnarray}
W(x,t) = \frac{1}{2} \sum_{l,n=-\infty}^{\infty} (-1)^{n\cdot l}
&& \left\{ \tilde{\varphi}_-(\kappa_n)
\, \varphi_+[2L\chi_{n,l}(x,t)] +
\tilde{\varphi}_-(-\kappa_n) \, \varphi_+[-2L\chi_{n,l}(x,t)]  \right.
\nonumber \\
&&- \left. \tilde{\varphi}_+(\kappa_n) 
\, \varphi_-[2L\chi_{n,l}(x,t)] -
\tilde{\varphi}_+(-\kappa_n) \, \varphi_-[-2L\chi_{n,l}(x,t)] 
\right\} \,.
\label{probability}
\end{eqnarray}
When we change in the second and forth term the summation over 
$n$ and $l$ by defining $n'\equiv-n$ and $l'\equiv -l$ and note from
the definition Eq.~(\ref{chi}) of $\chi_{n,l}$ the relation
\begin{equation}
\chi_{-n,-l}(x,t) = - \chi_{n,l}(-x,t)
\end{equation}
we arrive at the equivalent formulation
\begin{eqnarray}
W(x,t) = \frac{1}{2} \sum_{l,n=-\infty}^\infty (-1)^{n\cdot l}
&& \left\{ \tilde{\varphi}_-(\kappa_n) 
           \left[ \varphi_+ \left( 2L\chi_{n,l}(x,t) \right)
                 +\varphi_+ \left( 2L\chi_{n,l}(-x,t) \right)
           \right] \right. \nonumber \\
&& \left. - \tilde{\varphi}_+(\kappa_n)
           \left[ \varphi_- \left( 2L\chi_{n,l}(x,t) \right)
                 +\varphi_- \left( 2L\chi_{n,l}(-x,t) \right)
           \right] \right\}
\label{new}
\end{eqnarray}

This result is valid for any factorizable product $\varphi^*(x') \, 
\varphi(x'')$ of initial wave packets.
It brings out most clearly the way in which the initial conditions
determine the visibility of the intermode traces.
Indeed, according to Eq.~(\ref{new}) the probability distribution 
$W(x,t)$ to find the particle at time $t$ and position $x$
is a superposition of  
wave functions $\varphi_-$ and $\varphi_+$, evaluated along the 
spacetime lines $\chi_{n,l}(x,t)$ and $\chi_{n,l}(-x,t)$.
The slope $n^{-1}$ of these world lines is selected by the Fourier
transforms $\tilde{\varphi}_-$ and $\tilde{\varphi}_+$, which
represent the momentum distribution associated with the 
{\it factorized\/} wave packets.
Hence, the factorized wave functions, 
$\varphi_-$ and $\varphi_+$, and their Fourier transforms, 
$\tilde{\varphi}_-$ and $\tilde{\varphi}_+$ play the role
of weight factors for the spacetime lines $\chi_{n,l}(x,t)$ and
$\chi_{n,l}(-x,t)$.
\section{Example}
\label{s5}
We now illustrate the result, Eq.~(\ref{probability}),
for the probability distribution $W(x,t)$ by applying it to the
special case of a Gaussian initial wave packet.
Such a packet does indeed enjoy the factorization
property Eq.~(\ref{fact_property}).
However an arbitrary Gaussian does not satisfy the boundary conditions
enforced by the walls.
For this reason we now focus on a Gaussian wave packet that is initially
much narrower than the width of the box.

We consider a Gaussian wave packet
\begin{equation}
g(x) \equiv \left( \sqrt{\pi} \Delta x \right)^{-1/2}
\exp\left[ -\frac{1}{2} \left( \frac{x-\overline{x}}{\Delta x} \right)^2
\right]  \exp\left[ i \overline{p} (x-\overline{x})/\hbar \right]
\end{equation}
of width $\Delta x$ centered at $x = \overline{x}$ and moving with
average momentum $\overline{p}\equiv \hbar \overline{k}>0$.
Indeed this wave packet satisfies the factorization condition
\begin{equation}
g^*(x') \, g(x'') = \varphi_+(x'+x'') \, \varphi_-(x'-x'')
\label{factorization}
\end{equation}
where the new factorized wave functions 
\begin{equation}
\varphi_+(y) \equiv \left( \sqrt{\pi} \Delta x \right)^{-1/2}
\exp \left[ -\left(\frac{y/2-\overline{x}}{\Delta x} \right)^2 \right]
\end{equation}
and 
\begin{equation}
\varphi_-(y) \equiv \left( \sqrt{\pi} \Delta \right)^{-1/2}
\exp \left[ -\left(\frac{y}{2\Delta x} \right)^2 \right] \,
\exp(-i\overline{p} y / \hbar) 
\end{equation}
are also Gaussian wave packets.

After calculating the corresponding Fourier transforms and inserting
them into Eq.~(\ref{probability}) for the probability distribution
$W(x,t)$ to find the particle at time $t$ and position $x$, one 
arrives at the following expression
\begin{eqnarray}
W(x,t) = \frac{1}{2L} \sum_{l,n=-\infty}^{\infty} (-1)^{n\cdot l}
&& \left\{ \exp\left[-\left( \frac{L\chi_{n,l} - \overline{x}}{\Delta x}
\right)^2 \right] \, \exp\left[ - \left( 
\frac{\kappa_n-\overline{k}}{\Delta
\kappa} \right)^2\right] \right. \nonumber \\
&&+ \exp\left[-\left( \frac{L\chi_{n,l} + \overline{x}}{\Delta x}
\right)^2 \right] \, \exp\left[ - \left( 
\frac{\kappa_n+\overline{k}}{\Delta
\kappa} \right)^2\right] \nonumber \\
&&- \left. 2 \exp\left[ -\left( \frac{\kappa_n}{\Delta\kappa} 
    \right)^2 \right] \,
\exp\left[ -\left( \frac{L\chi_{n,l}}{\Delta x} \right)^2\right] \,
\cos\left[ 2( \overline{k} L \chi_{n,l} - \kappa_n \overline{x} )
\right] \right\} \,,
\label{prob_gauss}
\end{eqnarray}
where $\Delta \kappa \equiv 1 / \Delta x$.

We can identify in Eq.~(\ref{prob_gauss}) two different kinds of
structures among all the possible intermode traces.

We start our analysis from the
Gaussian $\exp[-(\kappa_n-\overline{k})^2/\Delta\kappa^2]$
which is centered around the average wave number $\overline{k}$.
It takes on dominant values only for
$|\kappa_n-\overline{k}|\sim \Delta\kappa$.
The other weight factor, 
$\exp[-(L\chi_{n,l}-\overline{x})^2/\Delta x^2)]$, selects the spacetime
lines starting at $t=0$ from the initial average position $\overline{x}$
of the particle in the box and from its images $\overline{x}+lL$,
periodically located along the $x$-axis.
The combined effect of these two factors is to restrict the
intermode traces to those connected with the classical motion.
They therefore have a slope $n^{-1}$ related to the average
momentum and emerge from the initial location of the particle in
in the potential well. 

A similar argument applies to the second term of Eq.~(\ref{prob_gauss}).
In this case the spacetime lines
$\chi_{n,l}(x,t)$ are characterized by a negative slope $-|n|^{-1}$ 
satisfying the condition $|\kappa_n + \overline{k}| \sim \Delta\kappa$.
They originate from $-\overline{x}+lL$.

The third term of Eq.~(\ref{prob_gauss}) has the form of an 
interference term, weighted by two Gaussians.
The Gaussian containing the quantized wave vector $\kappa_n$
is centered around zero. 
Therefore now the lines with small $n$, or
equivalently with large slope $n^{-1}$, are enhanced.
According to the weight factor $\exp[-(L\chi_{n,l}/\Delta x)^2]$
these structures emerge at the time $t=0$ from the walls of the
box, located at $x=0$ and $x=L$, and from their replicas periodically
placed at $x=lL$.
These are indeed the relevant intermode traces giving rise to
the regular pattern in the spacetime plot of the probability
distribution $W(x,t)$.

\begin{figure}[b]
\centerline{ \epsfig{file=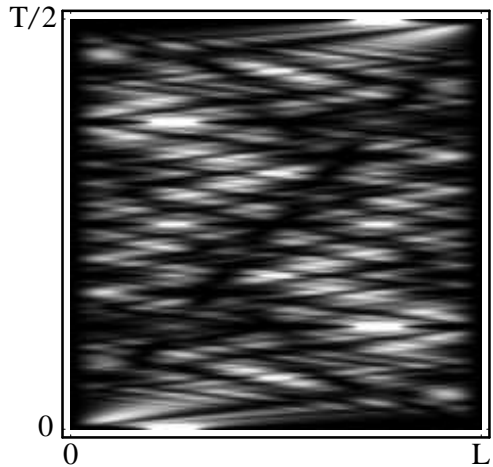,height=6.5cm} }
\caption{Density plot of the probability distribution $W(x,t)$ to
find the particle at time $t$ at position $x$.  
The initial wave packet is a Gaussian 
with width $\Delta x=L/20$ and average wave number $\overline{k}=0$
centered at $\overline{x}=L/4$.
Darker areas correspond to minima of probability, while brighter ones
represent maxima.
\label{zero} }
\end{figure}

\begin{figure}[htb]
\centerline{ \epsfig{file=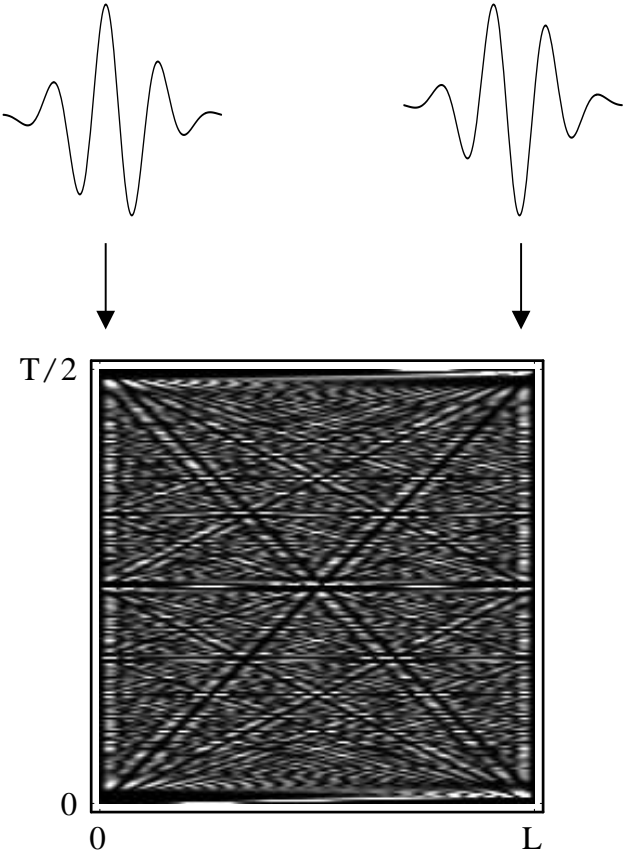,width=8.5cm} }
\caption{Density plot of the probability distribution
of the particle in a box as a function of position and time.  
The initial wave function is a Gaussian of width 
$\Delta x=L/20$ and average wave number $\overline{k}=20 \pi/L$
centered at $\overline{x}=L/4$.
On top we show a magnified view of the interference term in 
Eq.~(\protect\ref{prob_gauss}) 
for $n=1$, that is for the diagonal connecting the lower left
corner and the upper right corner, and for the other main 
diagonal corresponding to $n=-1$.  
Note that in the first case two ridges surround a canal,
while in the second case the situation is reversed and two canals
surround a ridge.
\label{combo} }
\end{figure}

The last factor, $\cos[2(\overline{k} L \chi_{n,l}
-\kappa_n \overline{x})]$, provides a modulation depending on
the initial conditions, that is average position and wave number.
In the extreme case of zero momentum, shown in Fig.~\ref{zero}, 
the cosine factor reduces to $\cos(n\pi\overline{x}/L)$.
Hence whenever its argument is equal to 
$\pm\pi/2\mbox{mod}(2\pi)$ the corresponding intermode traces 
are washed out.  
Indeed, in the example considered in Fig.~\ref{zero} the lines
$t/(T/2) = \pm 1/2 (x/L - l)$
are not present because the cosine factor $\cos(n\pi/4)$ vanishes
for $n=\pm 2$.
In the opposite limit of large average momentum, the cosine term
is responsible for the fine structure of the quantum carpet.
A comparison between Figs.~\ref{zero} and \ref{combo} shows that
by increasing the initial momentum, both canals and ridges develop
neighboring structures: Consider, for example, the lines running 
along the main diagonals.
In the case of non-vanishing initial wave numbers
they split into an alternating series of canals and ridges.
The detailed spatial behavior of such intermode traces is shown on top
of Fig.~\ref{combo}.
\section{Conclusions}
\label{s6}
We have studied the probability distribution of a particle in a
one-dimensional box as a function of position and time.
The corresponding density plot in spacetime is characterized by a 
regular pattern
of intermode traces, that is canals or ridges cutting through and
building onto an almost uniform background of probability.

In order to explain this phenomenon we have related the probability
distribution to the product of the initial wave packet and 
its complex conjugate via the kernel $K$.
This formulation allows us to distinguish between the intrinsic 
characteristics of the system and the influence of the specific
initial conditions.
 
Indeed we have found that the kernel, which consists of the product
of two Green's functions, contains already the spacetime structures.
It can be expressed as a superposition of $\delta$-functions whose
argument is a linear function of time and position.
The kernel is non-vanishing only along this set of infinitely many
lines with discrete steepness.
However, the initial conditions can either enhance or suppress the
visibility of such structures.
They therefore select the ones with the largest degeneracy from 
all possible intermode traces.

We have shown that an initial wave packet which satisfies a certain 
factorization
property is especially suitable to bring out the pattern
in the probability distribution. 
An example of such wave functions is the Gaussian wave packet, discussed
here. 
We emphasize however that the treatment presented here is more general.
It even allows us to investigate the case of a uniform initial 
distribution with discontinuities,
analyzed by Berry in the context of fractal properties of the spacetime
probability distribution.
\section*{Acknowledgement}
We express our gratitude to P.~J.~Bardroff, M.~V.~Berry, 
M.~Fontenelle, F.~Gro{\ss}mann,
M.~Hall, T.~Kiss, W.~E.~Lamb, Jr., 
K.~A.~H. van Leeuwen, C.~Leichtle, J.~Marklof,
M.~M.~Nieto, J.~M.~Rost, F.~Saif and P.~Stifter for many fruitful
discussions on this topic.
One of us (I.~M.) thanks the organizers of this conference for the
opportunity to report this work and for a most splendid meeting.
Two of us (I.~B.-B.) and (A.~E.~K.) thank the Humboldt Stiftung for
their support. 
\section*{Appendix: new representation of ${\cal D}$}
In this appendix we reexpress the function
\begin{equation}
{\cal D} (\eta ; \zeta) \equiv \sum_{m',m''=-\infty}^{\infty}
\exp[i \pi (m'+m'') \eta] \, 
\exp \left\{ -i\pi (m'-m'') \left[ \xi - (m'+m'')2\tau +\zeta
\right] \right\}
\label{D1}
\end{equation}
defining the kernel $K$, Eq.~(\ref{kernel1}), in terms of infinitely
many $\delta$-functions aligned along straight lines in spacetime.
Throughout this appendix we use the dimensionless position variables
$\xi$, $\eta$ and $\zeta$ normalized to the length of the box
and the dimensionless time $\tau$ scaled
with respect to the revival time $T$.

We note from the definition of ${\cal D}$ that only 
the sum $m'+m''$ and the difference $m'-m''$ of the summation indices
$m'$ and $m''$ occur.
This suggests to introduce new summation indices $m\equiv m'+m''$ and
$k\equiv m'-m''$.
However, this definition implies that $m'=(k+m)/2$ and $m''=(m-k)/2$.
Hence when one of the two new indices is odd and the other
is even the quantities $m'$ and $m''$ are not integer anymore.
Therefore, $m$ and $k$ must both be either even or odd leading to the
substitutions
\begin{equation}
2m = m' + m'' \quad \mbox{and} \quad 2k = m'-m''
\end{equation}
or
\begin{equation}
2m+1 = m' + m'' \quad \mbox{and} \quad 2k+1 = m'-m'' \,,
\end{equation}
and to the representation
\begin{eqnarray}
{\cal D} &=& \sum_{k,m=-\infty}^{\infty} e^{i\pi(2m)\eta} \,
\exp\left\{ -i\pi (2k) \left[ \xi - (2m)2\tau + \zeta \right]\right\}
\nonumber \\
&+& \sum_{k,m=-\infty}^{\infty} e^{i\pi(2m+1)\eta} \,
\exp \left\{-i\pi [ \xi -(2m+1) 2\tau + \zeta] \right\} \,
\exp\left\{ -i\pi (2k) \left[ \xi - (2m+1)2\tau +\zeta\right]\right\}\,.
\label{D2}
\end{eqnarray}
When we recall the relation
\begin{equation}
\sum_{k=-\infty}^{\infty} \exp(-2\pi i k\theta) = 
\sum_{l=-\infty}^{\infty} \delta(\theta-l)
\end{equation}
we can perform the summation over $k$ and arrive at
\begin{eqnarray}
{\cal D} &=& \sum_{l,m=-\infty}^{\infty} e^{i\pi(2m)\eta} \,
\delta[\xi -(2m)2\tau + \zeta -l] \nonumber \\
&+& \sum_{l,m=-\infty}^{\infty} e^{i\pi(2m+1)\eta} \,
\exp \left\{-i\pi [ \xi -(2m+1) 2\tau + \zeta] \right\} 
\, \delta[\xi -(2m+1)2\tau +\zeta -l]
\end{eqnarray}
which with the help of 
\begin{equation}
f(\xi) \, \delta(\xi) = f(0) \, \delta(\xi)
\end{equation}
simplifies to
\begin{eqnarray}
{\cal D} &=& \sum_{l,m=-\infty}^{\infty} e^{i\pi(2m)\eta} \,
\delta[\xi -(2m)2\tau + \zeta -l] \nonumber \\
&+& \sum_{l,m=-\infty}^{\infty} e^{i\pi(2m+1)\eta} \, e^{-i\pi l} \,
\delta[\xi -(2m+1)2\tau +\zeta -l] \,.
\end{eqnarray}
We can combine the sum over $(2m)$ and the one over $(2m+1)$ into
one sum over $n$ when we recall the relation
\begin{equation}
e^{-i\pi l} = (-1)^l 
\end{equation}
and note that 
\begin{equation}
(-1)^{n\cdot l} = \left\{ \begin{array}{rl}
1      & \mbox{for $n=2m$} \\
(-1)^l & \mbox{for $n=2m+1$} \,.
\end{array} \right.
\end{equation}
This yields the new representation
\begin{equation}
{\cal D}(\eta ; \zeta) =  \sum_{n,l=-\infty}^{\infty} (-1)^{n\cdot l}
\, e^{i \pi n \eta} \, \delta[\xi - n(2\tau) +\zeta -l]
\end{equation}
which is central to the analysis of the kernel $K$ in Sec.~\ref{s3}.
\end{document}